\documentclass[twocolumn]{revtex4}

\usepackage{graphicx}
\usepackage{dcolumn}
\usepackage{amsmath}
\usepackage{color}

\begin{document}  

\title{Enhancement of superconductivity due to kinetic-energy effect 
in the strongly correlated phase of the two-dimensional Hubbard model
}

\author{Takashi Yanagisawa}

\affiliation{Electronics and Photonics Research Institute,
Advanced Manufacturing Research Institute,
National Institute of Advanced Industrial Science and Technology
1-1-1 Umezono, Tsukuba, Ibaraki 305-8568, Japan
}

\begin{abstract}
We investigated kinetic properties of correlated pairing states 
in strongly correlated phase of the Hubbard model in two space dimensions.
We employ an optimization variational Monte Carlo method, where 
we use the improved wave function 
$\psi_{\lambda}= e^{-\lambda K}\psi_G$ for the Gutzwiller
wave function $\psi_G$ with $K$ being the kinetic part of the Hamiltonian. 
The Gutzwiller-BCS state is stabilized as the potential energy driven
superconductivity because the Coulomb interaction energy is
lowered while the kinetic energy increases in this state. 
In contrast,
we show that in the $\psi_{\lambda}$-BCS wave function 
$\psi_{\lambda-BCS}= e^{-\lambda K}P_G\psi_{BCS}$, the Coulomb 
energy increases and instead the kinetic energy is lowered
in the strongly correlated phase where the Coulomb repulsive interaction
$U$ is large.
The correlated superconducting state is realized as a kinetic
energy driven pairing state and
this indicates the enhancement of superconductivity due to kinetic-energy
effect.
\end{abstract}


\maketitle

\section{Introduction}

The mechanism and various mysterious properties of
cuprate superconductors have been intensively studied\cite{bed86}.
It is significant to clarify the mechanism of superconductivity and
understand the ground state phase diagram. 
The solution of mechanism of high-temperature superconductivity will
open a way to design new high-temperature superconductors.
The CuO$_2$ plane plays an essentially important role in 
cuprates\cite{mce03,hus03,web08,hyb89,esk89,mcm90,esk91}.
The basic model of the CuO$_2$ plane is the d-p model (or called the three-band
Hubbard model)\cite{eme87,hir89,sca91,ogu94,koi00,yan01,yan03,
yan09,web09,lau11,web14,ave13,ebr16,tam16}.
When we neglect oxygen orbitals in the d-p model, we have the one-band
Hubbard model.
We regard the Hubbard model as an effective model of the d-p model
where oxygen degrees of freedom are effectively taken into account
in the one-band model. 
The Hubbard model\cite{hub63,hub64,gut63} has been studied since it
certainly contains essential physics of cuprate high-temperature
superconductors\cite{zha97,zha97b,yan95b,nak97,yam98,koi99,yam11,
har09,yan13a,bul02,yok04,yok06,aim07,miy02,yan08,yan13,yan16,yan19,yan19b}.
The Hubbard model was introduced by Hubbard to understand the 
metal-insulator transition\cite{mot74}.
The Hubbard model contains fruitful physics although it is very simple.
It exhibits interesting physics regarding high-temperature cuprates.
For example, we can understand antiferromagnetic insulator,
superconductivity,
stripes\cite{tra96,suz98,yama98,ara99,moc00,wak00,bia96,mai10,mon12,bia13} and
inhomogeneous states\cite{hof02,wis08,han04,yan01b,yin14,yan21c} 
based on the Hubbard model.

A quantum variational Monte Carlo method is useful in the investigation of
the ground state property in a strongly correlated system.
We use the optimization variational Monte Carlo method 
where we use the wave functions with $e^{-\lambda K}$ operator
where $K$ stands for the kinetic part of the Hamiltonian\cite{yan16,yan19,yan98,yan14}.

There is a possibility that the kinetic energy plays an important role in
realizing high-temperature superconductivity.
This issue, kinetic energy driven mechanism, has
been addressed for
the Hubbard model\cite{mai04,oga06,gul12,toc16} and the t-J
model\cite{fen03,wro03,guo07}.
Although this mechanism is referred to as the kinetic energy driven mechanism,
the origin of superconductivity in the Hubbard model is the on-site
Coulomb repulsive interaction. 
We discuss the kinetic energy enhancement of superconductivity in this paper.
In the BCS theory, the superconducting (SC) condensation energy
comes from the attractive potential energy.
In the Gutzwiller-BCS wave function, the SC condensation energy also
mainly comes from the Coulomb potential energy.
The Coulomb interaction energy is reduced in
the SC state compared to that in the normal state, and thus the SC state becomes 
stabilized.  In contrast, the kinetic energy gain stabilizes the SC
state for the improved wave function.
This results in the enhancement of superconductivity as a kinetic-energy
effect.

The paper is organized as follows.
In section II we show the model Hamiltonian.  In section III we discuss
the improved wave functions that we use in this paper.
We show the correlated SC wave function in section IV.
In section V we show results for the kinetic energy in SC states
and discuss the kinetic energy enhanced superconductivity.
A summary is given in the last section.

\section{Optimization variational Monte Carlo method}
\subsection{Hamiltonian and optimized wave functions}

The Hubbard Hamiltonian is given by
\begin{equation}
H= \sum_{ij\sigma}t_{ij}c^{\dag}_{i\sigma}c_{j\sigma}
+U\sum_in_{i\uparrow}n_{i\downarrow}.
\end{equation}
The parameters in this model are given as follows.
$t_{ij}$ indicates the transfer integral where
$t_{ij}=-t$ when $i$ and $j$ are nearest-neighbor 
pairs $\langle ij\rangle$ and
$t_{ij}=-t'$ when $i$ and $j$ are next-nearest neighbor pairs.
$U$ indicates the on-site Coulomb energy.
$N$ denotes the number of lattice sites and $N_e$ shows that of electrons.
The energy is measured in units of $t$ throughout this paper.

We evaluate the expectation
values of physical properties by using a Monte Carlo procedure.
We start from the Gutzwiller function which is written as
\begin{equation}
\psi_G = P_G\psi_0,
\end{equation}
where $P_G$ represents the Gutzwiller operatora.  $P_G$ is given by
$P_G= \prod_j(1-(1-g)n_{j\uparrow}n_{j\downarrow})$ with the parameter $g$ 
in the range of $0\le g\le 1$.  $\psi_0$
indicates a one-particle state for which we take, for example, the Fermi sea, 
the BCS state and atiferromagnetically ordered state.

The Gutzwiller function is improved by correlation operators
to take into account electron correlations.
We employ the wave function given by\cite{yan16,ots92,yan98,yan99,eic07,bae09,bae11}
\begin{equation}
\psi_{\lambda}= e^{-\lambda K}\psi_G,
\label{wf1}
\end{equation}
where $K$ is the noninteracting part of the Hamiltonian given by
\begin{equation}
K= \sum_{ij\sigma}t_{ij}c^{\dag}_{i\sigma}c_{j\sigma}.
\end{equation}
$\lambda$ is a real parameter.
We use the auxiliary field method to calculate expectation values.

\subsection{Superconducting state with correlation}

The BCS wave function is
\begin{equation}
\psi_{BCS}= \prod_k(u_k+v_kc^{\dag}_{k\uparrow}c^{\dag}_{-k\downarrow})
|0\rangle,
\end{equation}
with coefficients $u_k$ and $v_k$ appearing in the ratio
$u_k/v_k=\Delta_k/(\xi_k+\sqrt{\xi_k^2+\Delta_k^2})$, where
$\Delta_k$ is the gap function and
$\xi_k=\epsilon_k-\mu$ is the dispersion relation.
We adopt the $d$-wave symmetry $\Delta_k= \Delta_{sc}(\cos k_x-\cos k_y)$.
The Gutzwiller-BCS state is
\begin{equation}
\psi_{G-BCS}=P_{N_e}P_G\psi_{BCS},
\end{equation}
where $P_{N_e}$ indicates the operator that extracts the state with $N_e$
electrons.  This wave function is referred to as the resonating-valence
bond (RVB) state in the literature\cite{and87}:  
$\psi_{RVB}= \psi_{G-BCS}$.

The improved correlated superconducting wave function is
\begin{equation}
\psi_{\lambda-BCS}= e^{-\lambda K}P_G\psi_{BCS}.
\end{equation}
This wave function is called the $\lambda$-BCS state in this paper.
In the formulation of $\psi_{\lambda}$, we use the
electron-hole transformation for down-spin electrons:
$d_k= c^{\dag}_{-k\downarrow}$, $d^{\dag}_k= c_{-k\downarrow}$,
and the operator for up-spin electrons remains the same:  $c_k= c_{k\uparrow}$.
In the real space we have $c_i=c_{i\uparrow}$ and $d_i=c^{\dag}_{i\downarrow}$.
The pair operator $c^{\dag}_{k\uparrow}c^{\dag}_{-k\downarrow}$
is transformed to $c^{\dag}_kd_k$.
We can use the auxiliary field method in a
Monte Carlo simulation\cite{yan07}.

\section{$e^{-\lambda K}$ and the renormalization group method}

Let us discuss on the role of $K$ in the wave function.
We write $\psi_0$ in the form
\begin{equation}
\psi_0 = \sum_j a^0_j\varphi^0_j.
\end{equation}
$\{\varphi^0_j\}$ denotes a set of basis functions
where $j$ represents the label for the
electron configuration.
$\psi_{\lambda}$ is given as
\begin{equation}
\psi_{\lambda}= \sum_j a^0_je^{-\lambda K}P_G\varphi^0_j. 
\end{equation}
$\psi_{\lambda}$ is written as 
\begin{equation}
\psi_{\lambda}= \sum_j a^{\lambda}_j\varphi_j, 
\end{equation}
where $\{\varphi_j\}$ is a set of basis states.
The set $\{\varphi_j\}$ may include $\{\varphi^0_j\}$
because some coefficients
$a^0_{\ell}$s may vanish accidentally in the non-interacting state.

We now show the ground state energy $E/N$ as well as the kinetic energy
$E_{kin}=\langle K\rangle$ and the Coulomb energy $E_U$ as functions
of $U$ in Fig. 1.
The kinetic energy part gives a large contribution to the ground-state
energy $E$ when $U$ is large.
When $U>10t$, $E_U$ for $\psi_{\lambda}$ almost agrees with
that for $\psi_G$.  The difference of $E_{kin}$ for $\psi_{\lambda}$
and $\psi_G$ increases when $U>10t$. 

Let us investigate the role of $K$ from the viewpoint of excitations in 
the momentum space.
The operator $e^{-\lambda K}$ controls the weights of excitation
modes in the Gutzwiller function $P_G\psi_0$.
It is seen that $e^{-\lambda K}$ suppresses high-energy excitations
since the eigenvalues of $K$ are large and then $e^{-\lambda K}$
becomes small.
This tells us that $e^{-\lambda K}$ plays a role of the projection
operator that projects out low lying excitation modes.
We point out that the role of $e^{-\lambda K}$ is similar to that of the
renormalization group procedure.
When the cutoff $\Lambda$ ($\sim$ the bandwidth) reduces to $\Lambda-d\Lambda$,
the states near the Fermi surface are magnified and their contribution
increases\cite{wil75}.
The increase of the parameter $\lambda$ corresponds to reducing
contributions from high-energy modes excited by
the Gutzwiller operator.

The effect of $e^{-\lambda K}$ is clearly reflected in the momentum
distribution function 
$n_{{\bf k}}=\langle c^{\dag}_{{\bf k}\sigma}c_{{\bf k}\sigma}\rangle$.
We show $n_{{\bf k}}$ in Fig. 2 where we put $U=10t$ and $N_e=88$
on a $10\times 10$ lattice.
$n_{{\bf k}}$ evaluated by using the Gutzwiller function presents
an unphysical behavior where $n_{{\bf k}}$ near the Fermi surface
is greater than that at other wave numbers. 
This shortcoming of the Gutzwiller function is remedied by
$e^{-\lambda K}$ in the improved wave function.

\begin{figure}[htbp]
\centering
\includegraphics[width=7.0cm]{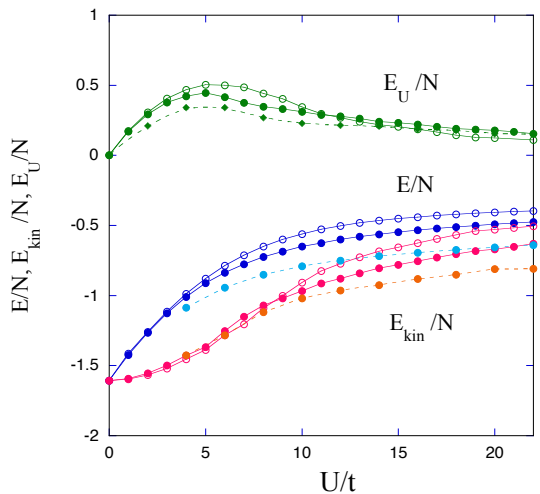}
\caption{
Ground-state energy $E/N$, kinetic energy $E_{kin}/N$ and the Coulomb
energy $E_U/N$ for $\psi_{\lambda}$ as a function of $U$
on a $10\times 10$ lattice. We use $N_e=88$ and $t'=0$ with the
periodic boundary condition in one direction and antiperiodic one in the
other direction.
The expectation values for the Gutzwiller function are also shown
by open circles.
The results $E/N$, $E_{kin}/N$ and $E_U/N$ for $N_e=80$ evaluated by 
$\psi_{\lambda}$ are
also shown by dashed lines.
}
\label{fig1}
\end{figure}

\begin{figure}[htbp]
\centering
\includegraphics[width=7.0cm]{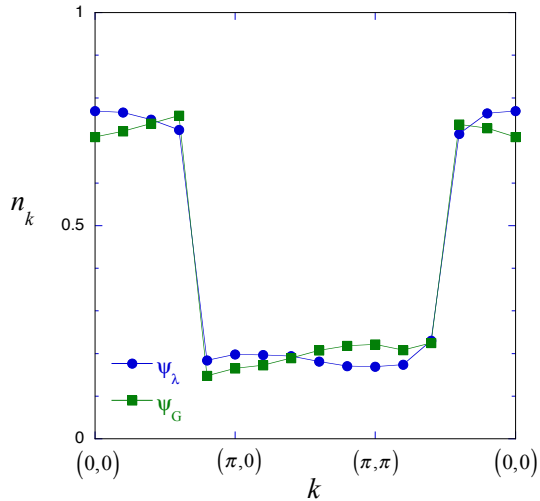}
\caption{
Momentum distribution function $n_k$ for $N_e=88$ and $U=10t$
on a $10\times 10$ lattice.
The results for $\psi_{\lambda}$ (circles) and $\psi_G$ (squares) are shown.
}
\label{fig2}
\end{figure}

\section{Kinetic energy enhancement of superconductivity}

\subsection{Why does the Gutzwiller-BCS state become stable?}

In the original BCS theory, the superconducting condensation
energy comes from the attractive potential interaction.
Let us examine the reason why the Gutzwiller-BCS state $P_G\psi_{BCS}$
becomes stable in the presence of the on-site Coulomb repulsive
interaction.
We show the kinetic energy $E_{kin}$ and the Coulomb energy $E_U$
in Fig. 3 and Fig. 4 for $P_G\psi_{BCS}$, respectively.
The Coulomb energy decreases as $\Delta_{sc}$ increases and
at the same time the kinetic energy increases.
The total ground energy $E$ has a minimum 
as shown in Fig. 5.
We can say that the Gutzwiller-BCS state belongs to the same
class of superconductivity in the sense that superconductivity is
induced by the potential energy.
This shows that the Gutzwiller-BCS state $\psi_{G-BCS}=P_G\psi_{BCS}$
is stabilized due to the reduction of the Coulomb potential energy
in a similar way to the BCS state, indicating that the
Gutzwiller-BCS superconductivity is a potential energy driven
superconductivity.

\begin{figure}
\centering
\includegraphics[width=7.0cm]{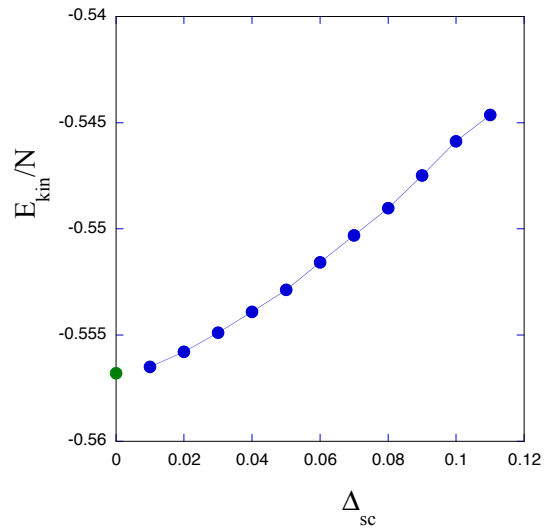}
\caption{
Kinetic energy $E_{kin}/N$ for the Gutzwiller function  as a 
function of the gap function 
$\Delta_{sc}$ for $U=18$ and $N_e=88$ on a $10\times 10$ lattice.
The extrapolated value for $\Delta_{sc}\rightarrow 0$ is shown on
the y-axis.
}
\label{fig3}
\end{figure}

\begin{figure}
\centering
\includegraphics[width=7.0cm]{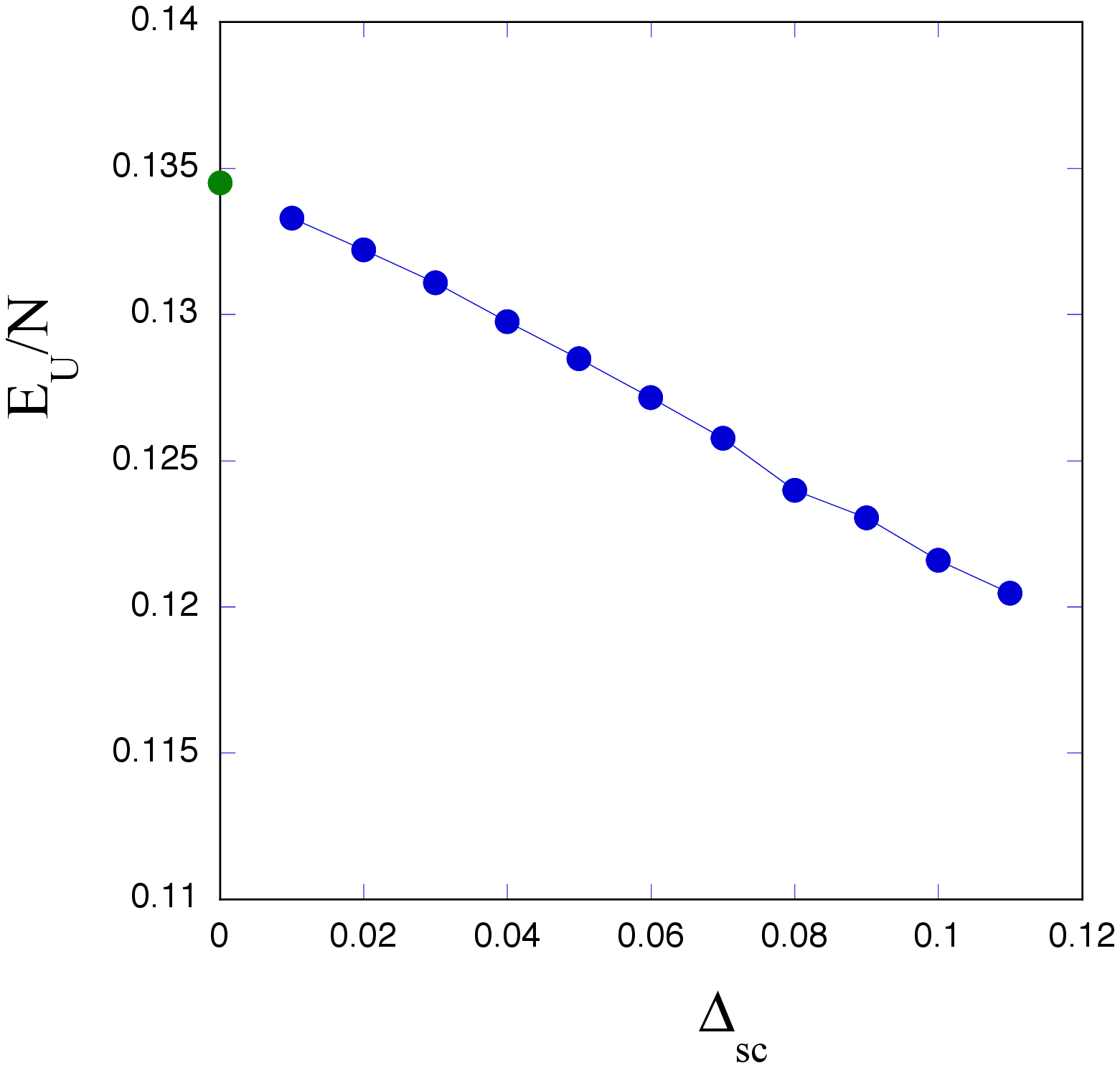}
\caption{
Coulomb energy $E_{U}/N$ for the Gutzwiller function as a function 
of the gap function 
$\Delta_{sc}$ for $U=18$ and $N_e=88$ on a $10\times 10$ lattice.
The extrapolated value for $\Delta_{sc}\rightarrow 0$ is shown on
the y-axis.
}
\label{fig4}
\end{figure}

\begin{figure}
\centering
\includegraphics[width=7.0cm]{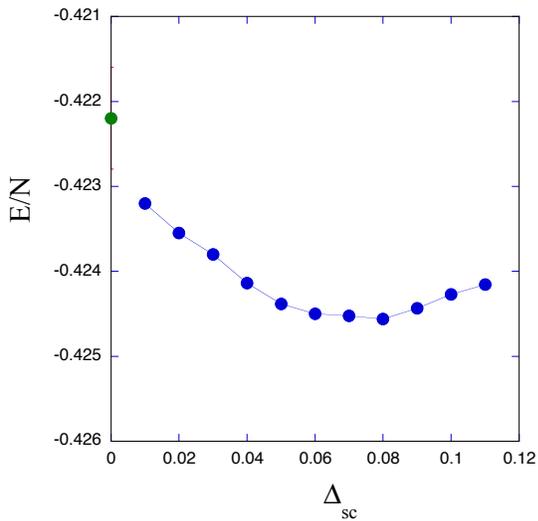}
\caption{
Ground-state energy $E/N$ for the Gutzwiller function as a function 
of the gap function 
$\Delta_{sc}$ for $U=18$ and $N_e=88$ on a $10\times 10$ lattice.
The circle on the y-axis  denotes the value extrapolated from
$E_{kin}$ and $E_U$ for $\Delta_{sc}\rightarrow 0$.
}
\label{fig5}
\end{figure}

\subsection{Why is the SC condensation energy so small?}

The $U$-part of the condensation energy $\Delta E_{U-sc}$ is clearly
proportional to $U$.  Then $\Delta E_{U-sc}$ ca be large when $U$ is
large and we can expect high-temperature superconductivity.
The SC condensation energy $\Delta E_{sc}$ is, however, 
very small compared to the transfer $t$.
$\Delta E_{sc}$ becomes very small due to the offset of $\Delta E_{kin-sc}$
and $\Delta E_{U-sc}$.
As a result the SC transition temperature $T_c$ is very much lower
than we expect.
We can say that $\Delta E_{sc}$ is determined by the competition between
kinetic energy effect and interaction effect.
Two competitions occur where one is the competition between
superconductivity and antiferromagnetism and then the other occurs between
kinetic energy and interaction energy.    
The superconducting transition occurs as a result of two competitions.

\subsection{How does the $\psi_{\lambda-BCS}$ state become
stable?}

We turn to the improved off-diagonal function $\psi_{\lambda}$.
We estimate the kinetic energy in the superconducting state
$\psi_{\lambda-BCS}$\cite{yan21}.
We define the SC condensation energy $\Delta E_{sc}$ as a sum of
two contributions $\Delta E_{kin-sc}$ and
$\Delta E_{U-sc}$:
\begin{eqnarray}
\Delta E_{sc} &=& E(\Delta=0)-E(\Delta=\Delta_{{\rm opt}}),\\
\Delta E_{kin-sc} &=& E_{kin}(\Delta=0)-E_{kin}(\Delta=\Delta_{{\rm opt}}),\\
\Delta E_{U-sc} &=& E_{U}(\Delta=0)-E_{U}(\Delta=\Delta_{{\rm opt}}),
\end{eqnarray}
where $\Delta=\Delta_{sc}$ is the SC order parameter and $\Delta_{{\rm opt}}$
is the optimized value which gives the energy minimum.  We have
\begin{equation}
\Delta E_{sc}= \Delta E_{kin-sc}+\Delta E_{U-sc}.
\end{equation}
The kinetic energy in $\psi_{\lambda-BCS}$ is lower than the kinetic energy
in the normal state $\psi_{\lambda}$, which is shown in Fig. 6.
The Coulomb energy expectation value increases as $\Delta_{sc}$
increases as shown in Fig. 7.
The results show
\begin{equation}
\Delta E_{kin-sc}>0,~~~~ \Delta E_{U-sc}<0,
\end{equation}
for $\psi_{\lambda-BCS}$ with $U=18t$ and the hole density $x=0.12$.
Then the ground state becomes superconducting 
as shown in Fig. 8 where the ground state energy $E$ is shown as a 
function of $\Delta_{sc}$.
This is in contrast to the Gutzwiller-BCS state and original
BCS state for which $\Delta E_{kin-sc}<0$ and $\Delta E_{U-sc}>0$. 
We summarize this in Table. 1.

\begin{table}
\caption{Variations of the kinetic and potential energies in the
superconducting state compared to the normal state.
$T$ and $V$ denote the kinetic energy and potential energy,
respectively.
}
\begin{center}
\begin{tabular}{cccc}
\hline
State & T  &  V  &  \\
\hline
BCS  &  $\Delta T>0$  &  $\Delta V <0$  & weak coupling SC \\
Gutzwiller-BCS  &  $\Delta T>0$  &  $\Delta V <0$  & weakly correlated SC \\
$\lambda$-BCS  &  $\Delta T<0$  &  $\Delta V >0$  & strongly correlated SC \\
\hline
\end{tabular}
\end{center}
\end{table}

\subsection{Kinetic energy enhancement of superconductivity
}

We define the difference of the kinetic energy as
\begin{equation}
\Delta E_{kin}= E_{kin}(\psi_G)-E_{kin}(\psi_{\lambda}),
\end{equation}
where $E_{kin}(\psi_G)$ and $E_{kin}(\psi_{\lambda})$ indicate
the kinetic energy for $\psi_G$ and $\psi_{\lambda}$,
We can write $\Delta E_{kin}= E_{kin}(\lambda=0)-E_{kin}(\lambda)$
for the optimized value of $\lambda$. 
$\Delta E_{kin}$ has the close relation
with the SC condensation energy $\Delta E_{sc}$ and its kinetic
part $\Delta E_{kin-sc}$.

We show $\Delta E_{kin}/N$ in Fig. 9 for $x=0.12$ where
$x$ is the hole doping rate.
The Coulomb energy $E_U/N$ and the superconducting
condensation energy $\Delta E_{sc}/N$ are also shown in Fig. 9.
$\Delta E_{kin}$ begins to increase after the Coulomb energy
$E_U$ reaches the peak when $U\approx 8t$.
The y axis on the right shows $\Delta E_{kin-sc}/N$ in Fig. 9.
$\Delta E_{kin-sc}$ shows a similar behavior
to $\Delta E_{kin}$.
$\Delta E_{kin-sc}$ may change sign as a function of $U$, which is
consistent with the analysis for 
Bi$_2$Sr$_2$CaCu$_2$O$_{8+\delta}$\cite{deu06}.
This shows the kinetic energy enhancement of superconductivity.

We compare the kinetic energy difference
$\Delta E_{kin}/N$ for $x=0.12$ (the electron density $n=0.88$)
and $x=0.20$ ($n=0.80$) in Fig. 10.  $\Delta E_{kin}/N$ decreases when
the hole density increases.  

\begin{figure}
\centering
\includegraphics[width=7.0cm]{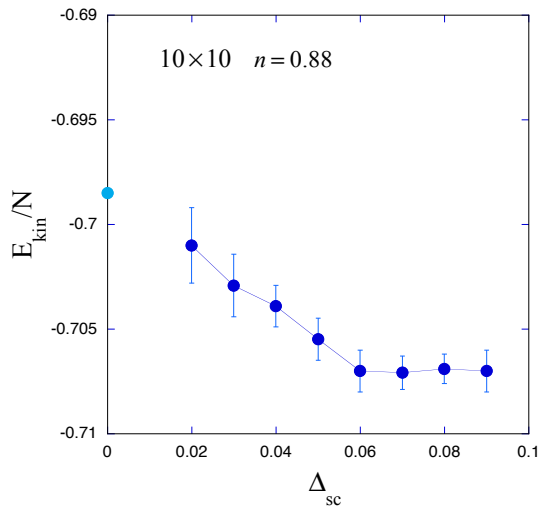}
\caption{
Kinetic energy $E_{kin}/N$ for $\psi_{\lambda-BCS}$ as a function of 
the gap function 
$\Delta_{sc}$ for $U=18$ and $N_e=88$ on a $10\times 10$ lattice.
The circle for $\Delta_{sc}=0$ denotes the extrapolated value.
}
\label{fig6}
\end{figure}

\begin{figure}
\centering
\includegraphics[width=7.0cm]{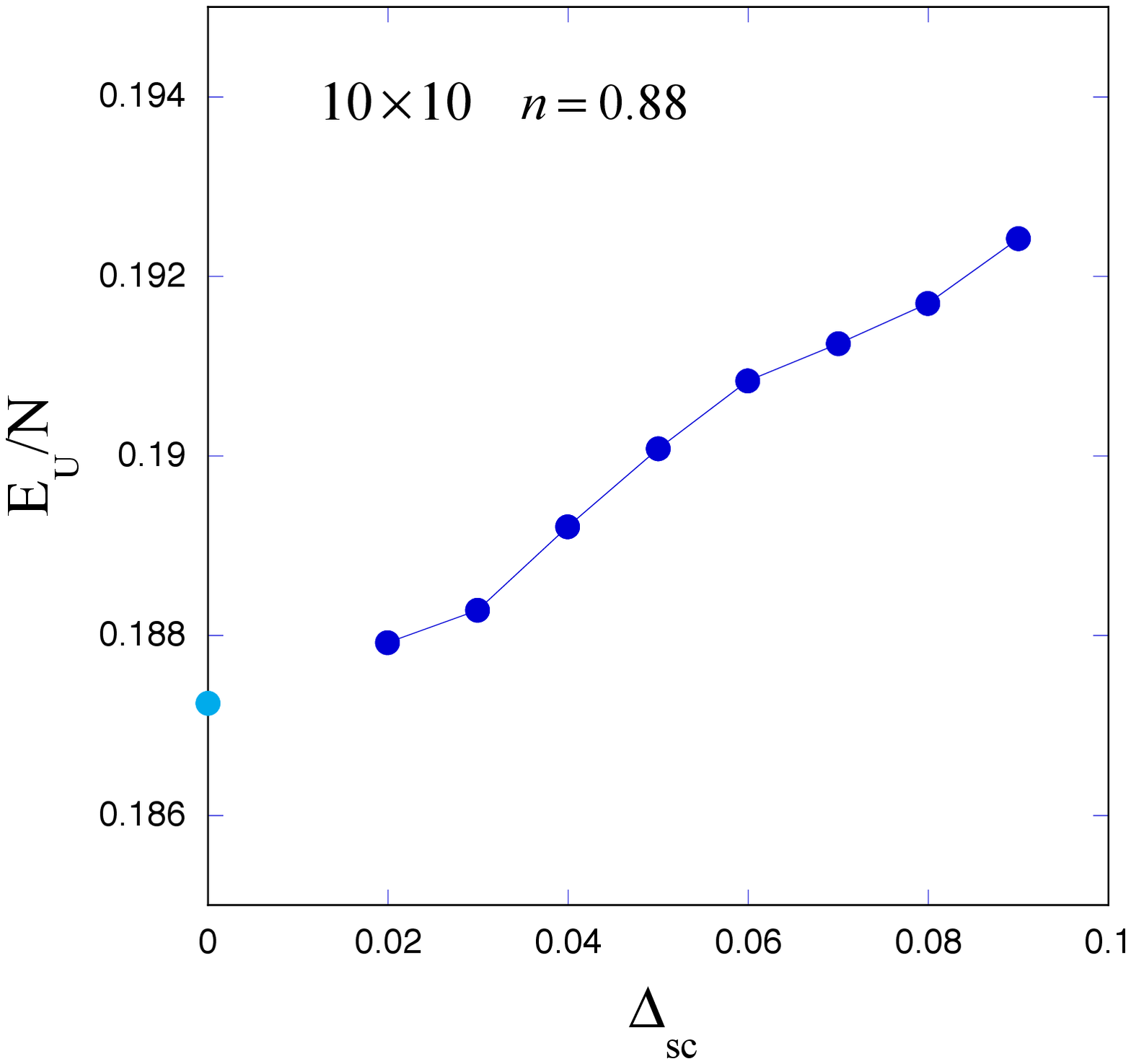}
\caption{
Coulomb energy $E_{U}/N$ for $\psi_{\lambda-BCS}$ as a function of 
the gap function 
$\Delta_{sc}$ for $U=18$ and $N_e=88$ on a $10\times 10$ lattice.
The circle for $\Delta_{sc}=0$ denotes the extrapolated value.
}
\label{fig7}
\end{figure}

\begin{figure}
\centering
\includegraphics[width=7.0cm]{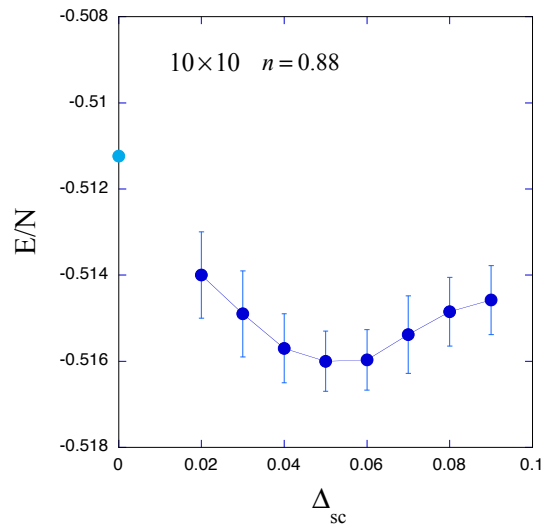}
\caption{
Ground-state energy $E/N$ for $\psi_{\lambda-BCS}$ as a function of 
the gap function 
$\Delta_{sc}$ for $U=18$ and $N_e=88$ on a $10\times 10$ lattice.
The circle for $\Delta_{sc}=0$ denotes the extrapolated value 
which agrees with the value obtained by evaluations for the normal state
wave function within statistical error.
}
\label{fig8}
\end{figure}

\begin{figure}
\centering
\includegraphics[width=7.5cm]{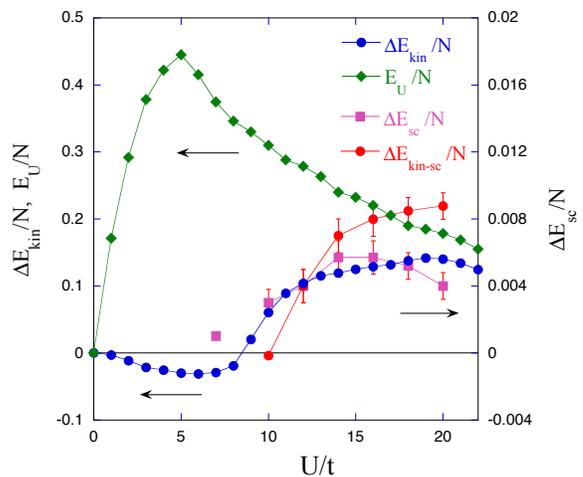}
\caption{
Kinetic-energy difference $\Delta E_{kin}/N$ and the kinetic-energy
gain $\Delta E_{kin-sc}/N$ in the superconducting state $\psi_{\lambda-BCS}$
as a function of $U$ on a $10\times 10$ lattice
where $N_e=88$ and $t'=0$.
The Coulomb energy $E_U/N$ for $\psi_{\lambda}$ and the condensation
energy $\Delta E_{SC}$ are also shown. 
We use the same periodic and antiperiodic boundary conditions as in Fig. 1.
The y axis on the right shows the superconducting condensation
energy $\Delta E_{sc}$ and the kinetic condensation energy 
$\Delta E_{kin-sc}/N$ for $\psi_{\lambda}$.
}
\label{fig9}
\end{figure}

\begin{figure}
\centering
\includegraphics[width=7.5cm]{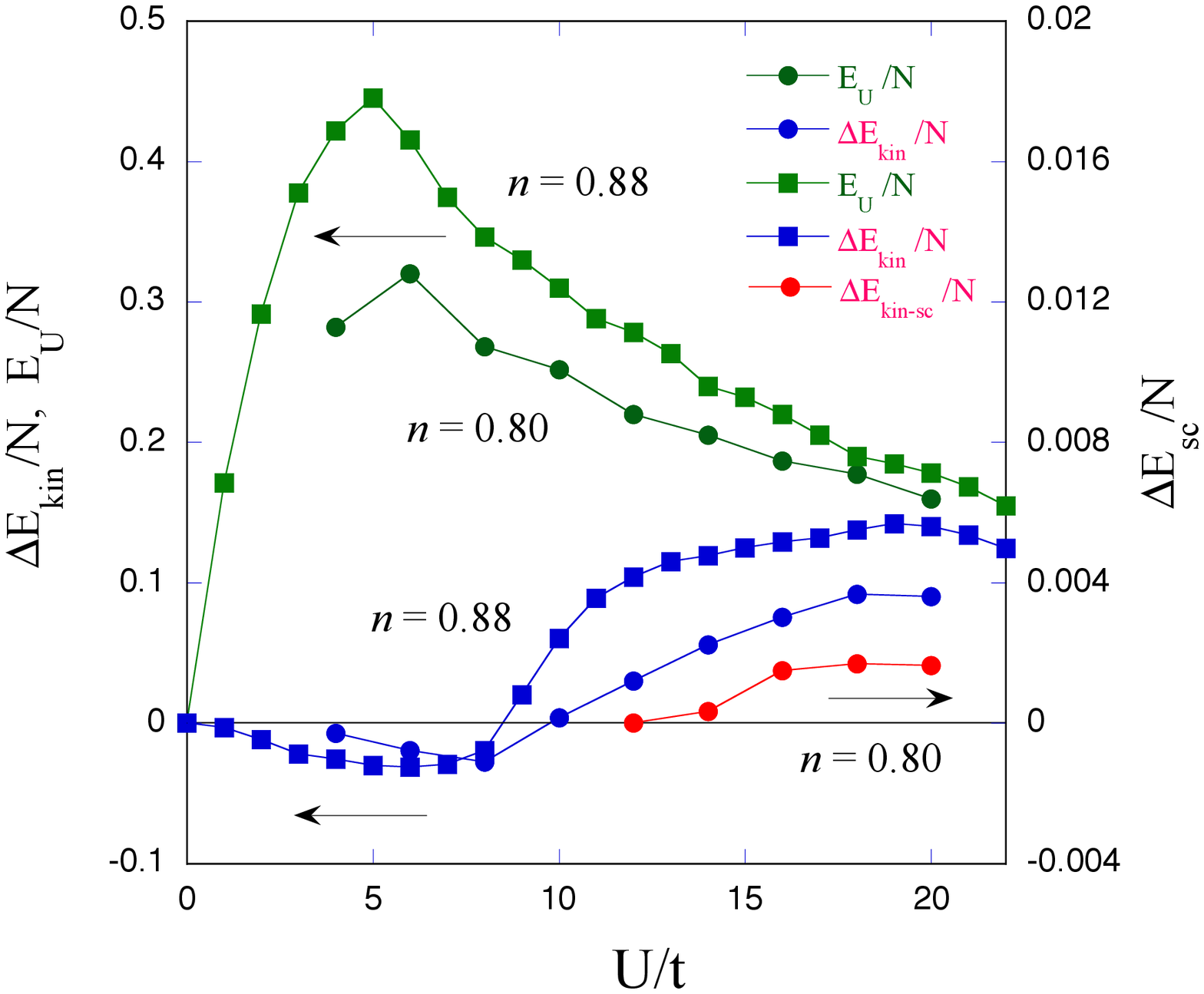}
\caption{
The Coulomb energy $E_U/N$,
Kinetic-energy difference $\Delta E_{kin}/N$ and the kinetic-energy
gain $\Delta E_{kin-sc}/N$
as a function of $U$ on a $10\times 10$ lattice where
$N_e=80$ ($n=0.80$).
$\Delta E_{kin}/N$ and $E_U/N$ for $n=0.88$ are also shown
for comparison.
We use the same boundary conditions as in Fig. 1.
}
\label{fig10}
\end{figure}

\section{Summary}

We have investigated the electronic properties of the two-dimensional 
Hubbard model by using the wave function with $e^{-\lambda K}$
correlation operator.
The operator $e^{-\lambda K}$ plays a role that is similar to the
renormalization group method.
$e^{-\lambda K}$ suppresses high-energy modes and  would project out
low lying modes near the Fermi surface.  This is typically shown
in the behavior of momentum distribution function $n_{{\bf k}}$.

We examined the kinetic energy effect in SC states.
The Gutzwiller-BCS state is the potential energy driven SC
state, because the SC condensation energy comes from the
Coulomb interaction energy.
This is the same as the original BCS state where the superconductivity
appears due to the attractive interaction.
We evaluated the kinetic energy in the improved SC state ($\psi_{\lambda-BCS}$)
to find that the kinetic energy gain stabilizes SC state while the
expectation value of Coulomb interaction energy increases.
This indicates that superconductivity is enhanced due to the kinetic
energy effect.
 
The kinetic energy difference $\Delta E_{kin}=E_{kin}(\psi_G)-E_{kin}(\psi_{\lambda})$
changes sign at $U/t\sim 9$ and
increases for $U>10t$.
$\Delta E_{kin-sc}$ in the SC state behaves like
$\Delta E_{kin}$ for $U>10t$, showing
a correlation between $\Delta E_{kin-sc}$ and $\Delta E_{kin}$.
This indicates the kinetic energy enhancement of superconductivity
in the strongly correlated phase.
The kinetic energy enhanced mechanism of superconductivity may be
different from the conventional mechanism of weak coupling superconductivity.

We here give a discussion on recent results by quantum Monte Carlo 
method\cite{qin20}, where the ground state of the 2D Hubbard model
is investigated for $U=8$ and the hole density $x=1/8$.
An antiferromagnetic correlation is large and the uniform SC state is
not stable for this set of parameters.  
We should examine the large-$U$ case where $U$ is greater than the
bandwidth so that an antiferromagnetic correlation is suppressed.
The striped state may be realized just at $x=1/8$.
In the striped state the paired holes mainly exist along stripes.
An inhomogeneous SC state will be realized and a pair correlation
function may be anisotropic.

We lastly discuss on the improvement of the wave function. 
The importance of the exponential factor $e^{-\lambda K}$ is clear
it is of course necessary to improve the wave function further
by multiplying $P_G$ and $e^{-\lambda K}$ again.
The improved wave function is written as
$\psi_m= e^{-\lambda_m K}P_G \cdots e^{-\lambda_1 K}P_G\psi_0$.
$\psi_m$ approaches the exact ground-state wave function as $m$
increases.  We believe that we obtain qualitatively the same result
for further improved functions because we obtained the finite SC
condensation energy in the limit $m\rightarrow\infty$\cite{yan99}.

A part of computations was supported by the Supercomputer
Center of the Institute for Solid State Physics, the University of
Tokyo and the Supercomputer system Yukawa-21 of the Yukawa Institute
for Theoretical Physics, Kyoto University.
This work was supported by a Grant-in-Aid for Scientific
Research from the Ministry of Education, Culture, Sports, Science and
Technology of Japan (Grant No. 17K05559).

\end{document}